\documentclass[%
 aip,
 amsmath,amssymb,
 reprint,%
]{revtex4-1}

\usepackage{graphicx}
\usepackage{dcolumn}
\usepackage{bm}

\usepackage[utf8]{inputenc}
\usepackage[T1]{fontenc}
\usepackage{mathptmx}
\usepackage{etoolbox}
\usepackage{ulem}

\usepackage{xcolor}
\makeatletter
\def\@email#1#2{%
 \endgroup
 \patchcmd{\titleblock@produce}
  {\frontmatter@RRAPformat}
  {\frontmatter@RRAPformat{\produce@RRAP{*#1\href{mailto:#2}{#2}}}\frontmatter@RRAPformat}
  {}{}
}%
\makeatother
\begin{document}

\preprint{AIP/123-QED}

\title[Sample title]{Enhanced hot electron generation from liquid jets in moderate intensity laser-plasma interactions }

\author{Ratul Sabui$^{1,2}$, S.V. Rahul$^1$, Angana Mondal$^1$, Archit Bhardwaj$^1$, Ram Gopal$^1$, Vandana Sharma$^2$, M. Krishnamurthy$^1*$}
 \affiliation{ 
$^1$Tata Institute of Fundamental Research, Hyderabad, India\\$^2$Indian Institute of Technology, Hyderabad, India
}%

\date{\today}

\begin{abstract}
We report the generation of MeV temperature electrons using sub-terawatt laser systems with a liquid methanol jet as a target. Remarkably, even at laser intensities of $\sim 10^{16} W/cm^2$, liquid cylindrical (2D) 15 $\mu m$ methanol jets produce electrons with temperatures of $\sim1$ MeV.  Hot electron emission characteristics are strikingly similar to those observed in spherical microdroplet (3D) targets. 
These results  validate that modeling such experiments using 2D PIC simulation is not a compromising approximation. This work further simplifies the experimental complexities towards a multi-KHz highly regenerative source of directed multi-MeV electron (and associated x-ray and ion) generation, demanding laser intensities 100x lower than conventional laser plasma sources. Increased source energy and pointing stability are crucial for  imaging or radiographic applications from such sources.

\end{abstract}
\maketitle

\section{\label{sec:level1}Introduction}

While the advent of high-intensity lasers has allowed the realization of astronomical phenomena in laboratory spaces, it has also opened up the possibility of having multimodal(electrons, hard x-rays, ions and neutrals) tabletop particle accelerators. Though various schemes(experimental and simulation) have exploited the ability of plasma to couple EM energy to charged particles, relativistic electrons with energies above ($\gamma - 1$) could only be achieved\cite{HuiChen2009,Beg1997} at intensities well above $10^{18} W/cm^2$.  Intensities of similar scale, as of yet, have scarcely been realized in high repetition rate systems. The complexity of producing and maintaining such high-intensity laser systems also pose an obstacle to most applications. Most practical applications demand high acceleration gradients and repeatably regenerating plasma targets, thus creating a niche space for studying novel targets in such systems. Size-limited target geometries have been studied extensively for their regenerative nature and higher absorption energies. While larger targets (>>100$\lambda_0$) allow for the redistribution of energy and subsequent thermalization, smaller targets allow for a higher energy density. Gas phase nanoclusters have shown almost 90\% absorption rates, emitting MeV ions and keV electrons\cite{ Kumarappan2003, Kumarappan2001, Jha2008} at $10^{16} W/cm^2$. Though foil targets are easier to generate and control than clusters, they show very low conversion at similar intensities, with particle energies confined below 100keV. This draws greater attention to mesoscopic targets, i.e., targets having sizes up to a few times the laser wavelength $\lambda_0$. To an extent, these targets embrace the mass-limited nature of nanoclusters, promising a more significant deposition of laser energy within the low target mass. Apart from that, these targets can be produced in large quantities, ensuring cost-effective repeatability. \par
In our experiments with spherical microscopic droplets(15$\mu m$ radius), it was realized that the introduction of a low-intensity replica of the main pulse 4ns ahead, results in a 20$\times$ enhancement in the accelerated electron temperatures. The findings were backed by 2D PIC simulations, with the size-limited spherical nature being held responsible for the enhancement. Computational facilities often limit users to 2D simulations over a more realistic 3D calculation. In the case presented here, the validity of a 2D simulation relies on its applicability to structures which are circular and size-limited in one plane, regardless of the geometry in the other plane. 
This paper explores the possibility of exploiting the 2 dimensional size-limited nature of liquid jets and provides a direct comparison to a 3 dimensional sphere, with electron acceleration being the major focus. We demonstrate experimentally that the cylindrical liquid methanol jet, when subjected to a pre-pulse induced modification can generate electrons of upto 4MeV, surpassing the existing scaling laws by almost 2 orders of magnitude. Thus we present a novel regenerative directional microscopic source of MeV electrons and x-rays that can operate on moderate-intensity multi-kHz Laser systems. Also, we establish the fact that the underlying mechanism predicted by our model is valid for plasma targets that are circular and size limited in just one plane. \par


\section{\label{sec:level1}The Experiment}

\begin{figure}[!ht]
\centering
  \includegraphics[width= \linewidth]{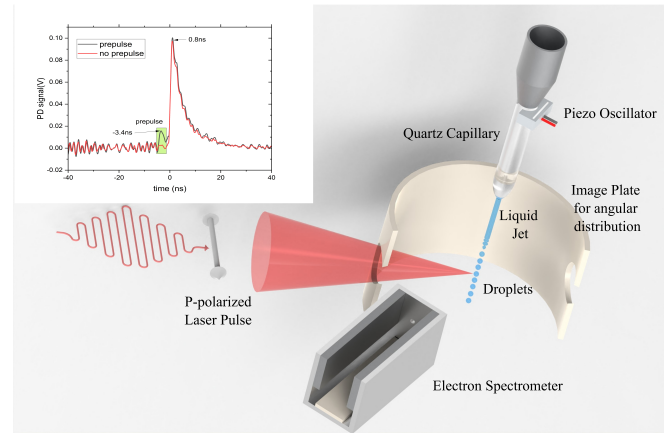}
    \caption{Experimental setup used in the study. The inset shows the laser temporal structure as detected on a fast PD. The pre-pulse is visible in this plot ahead of the main pulse. 
    }
    \label{expSet}
\end{figure}

The study involves a Ti-Sapphire based Laser system generating 4mJ 800nm pulses at 1kHz. The 35fs pulses were focused to a peak intensity of $4\times 10^{16} W/cm^2$ on the target. A pre-pulse, which is a  replica of the main pulse, arrives 4ns ahead to distort the sphericity of the plasma target. The prepulse has a peak intensity $<10\%$ of the main pulse. Methanol, acting as the target,  is delivered as a 15$\mu m$ diameter cylindrical jet at high pressure(5-10 Bar) through a quartz capillary. The jet spontaneously breaks into droplets on traversing over 1 cm in vacuum. The laser can be focused either on the cylindrical jet or on the spherical droplet depending upon the choice of point of interaction from the capillary. Thus the droplet/jet first interacts with the pre-pulse, evolves for 4ns and then interacts with the main pulse. The study compares jets and droplets of similar sizes, using the same target delivery system for both. A triggered piezo oscillator attached to the capillary ensures temporal synchronization of the laser and the droplet. Also the external perturbation guarantees the consistent generation of equi-sized droplets. BAS-MS image plates from Fujifilm are used as electron detectors, with the energy resolution provided by a magnetic spectrometer setup. The setup has been shown in Figure \ref{expSet}. The experiment is conducted in low vacuum conditions, with the evaporating liquid ensuring a chamber vacuum below 1mBar. \par

\begin{figure}[!ht]
\centering
  \includegraphics[width= 0.7\linewidth]{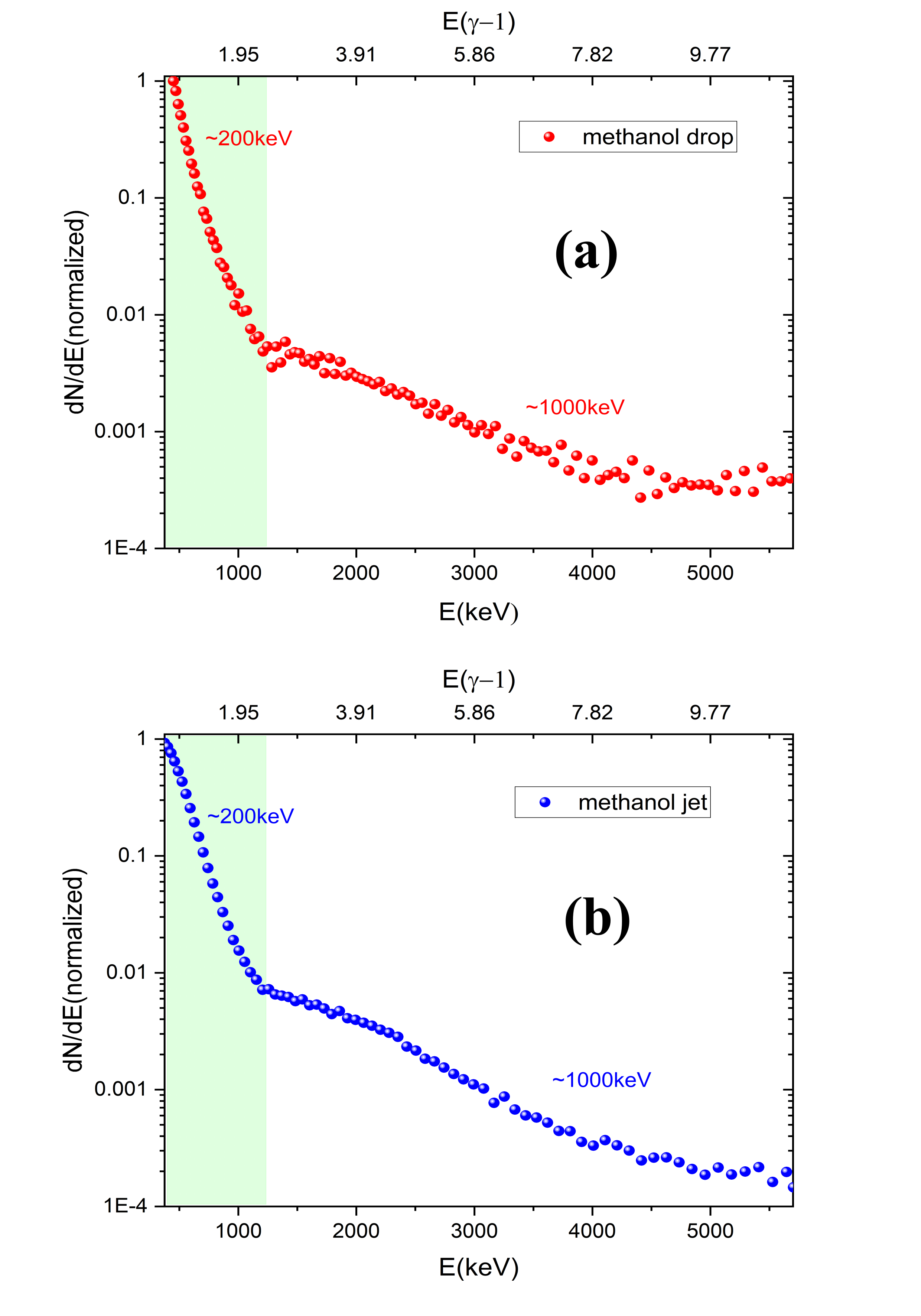}
    \caption{(a) and (b) show the electron energy spectrum measured when intense laser pulses at peak intensities of 4 $\times$10$^{16}$W cm $^{-2}$ are used on either 3D spherical microdroplets or 2D cylindrical jetc, resepectively, under other wise identical experimental conditions.  
    }
    \label{specComp}
\end{figure}

Figure \ref{specComp} shows the electron spectra collected from the interactions with the liquid droplets and the liquid jet. Prominent 200keV and 1000keV temperatures can be observed in both spectra, with the gradual high energy cutoff being close to 4500keV. In terms of high energy electrons, there is no significant variation between the two targets, suggesting the possibility of using the jet as a replacement source for droplets. Utilizing the liquid jets as an alternate source eliminates complications related to temporal synchronization in target generation. Detection of 1MeV temperature requires long image plate exposures, integrating electron counts over time. A comparison of the lower energy part of the spectrum has not been presented as a part of this study due to its dependence on multiple other factors. The surrounding gas environment has been observed to play a role in the emission of low energy electrons. This dependence shall be presented under a separate purview. The spectral window is chosen to compare the two major hot electron temperature components, about 200 keV and 1 MeV generated from both the targets.

\par
Focusing errors on moving targets occur due to the need for precise triggering of droplets, essential for proper laser interaction. The jet breaks into droplets due to the surface tension and the periodic instability generated by the perturbing acoustic wave. However, the time to break the stream to a drop for each wave period is statistical. This generates a positional jitter in the stream. Even a slight electronic timing jitter can lead to improper focusing, affecting electron and x-ray emission over time. The interaction position relative to the quartz capillary tip is crucial, with droplets(> 1cm) requiring longer travel distances than jets(<1mm). Droplet size and shape stability are observed approximately within 1cm from the tip, the distance varying with tip structure and pressure. Droplets are more susceptible to spatial variations due to the longer travel distance and lower inertia compared to jets. Minor spatial jitter at the source gets amplified over distance. While thicker jets offer increased stability, they come with their own set of drawbacks. Our study focuses on the generation of high energy electrons from the jets, avoiding a detailed study of multiple parameters needed to bring uniformity in the low energy counts. The utilization of the interaction as a possible electron source for radiography was verified for the cylindrical jet system(details in supplementary). Electron radiography performed on a $\mu m$ sized mesh revealed a resolution below 20$\mu m$ which also factors in the spatial jitter of the cylindrical jet. \par  

\begin{figure}[!ht]
\centering
  \includegraphics[width= 0.8\linewidth]{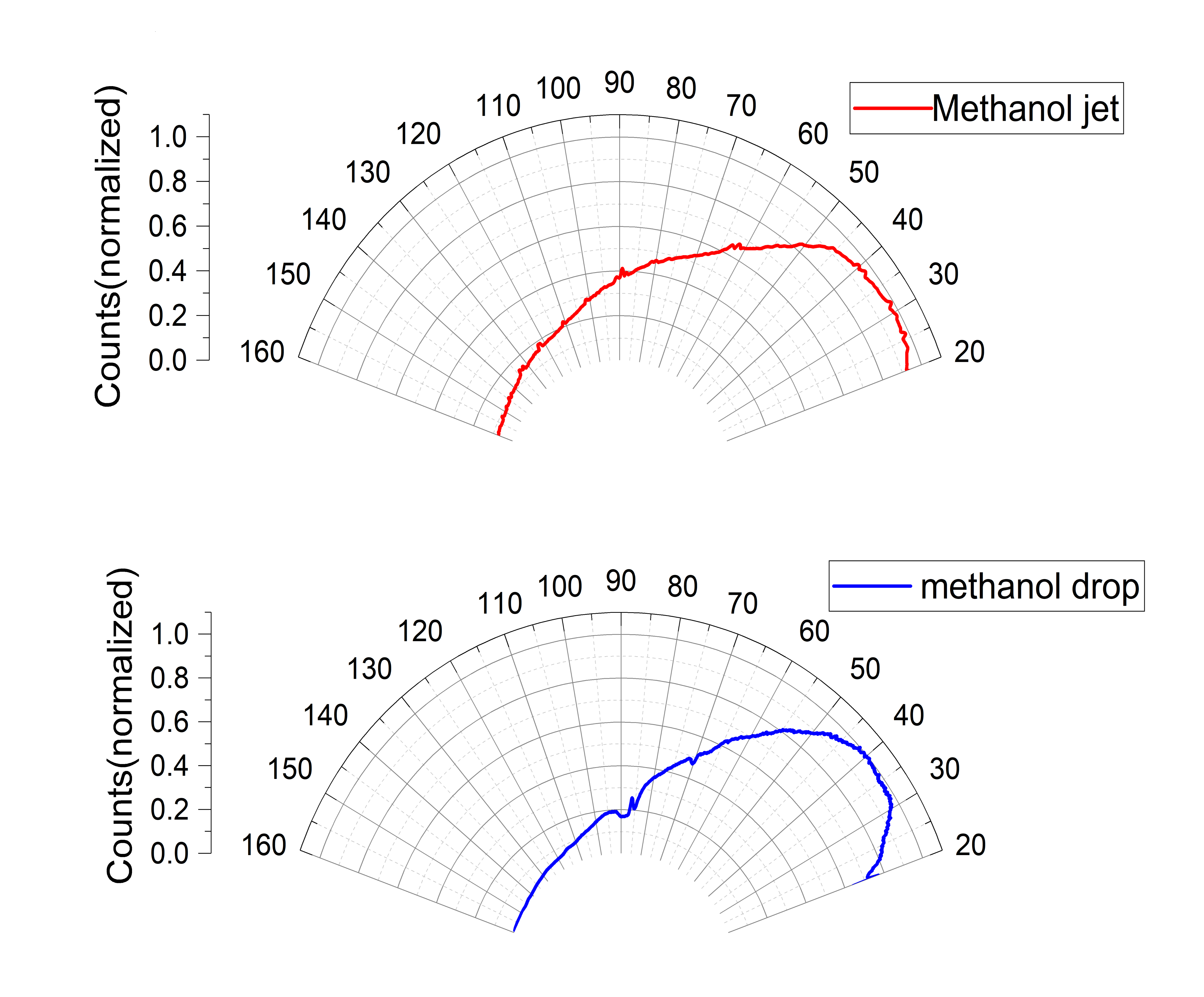}
    \caption{ (a) and (b) show the angular distribution of electron with more 100 keV energy,   measured when intense laser pulses at peak intensities of 4 $\times$10$^{16}$W cm $^{-2}$ are used on either 3D spherical microdroplets or  2D cylindrical jets, respectively, under other wise identical experimental conditions.
     The laser $\vec{k}$ originates at 0$^0$ and points towards 180$^0$  
    }
    \label{angDist}
\end{figure}

Apart from the enhancement observed in the electron energies, the directionality of the emission was closely replicated in both the droplet and the jet targets. 
Figure \ref{angDist} shows the angular distribution of electrons as observed on an image plate placed around the interaction point. Both the targets show maximal emission symmetric to the –$\vec{k}$ direction where $\vec{k}$ is the laser vector. The image plate was covered with 120 microns of aluminium to filter out electrons below 130keV, allowing only the higher energy electrons to reach the image plate. The emission angles in both cases peak between 20 and 45 degrees to the –$\vec{k}$ vector in the plane of polarization. As observed in our previous publications\cite{Mondal2024, Yembadi2024}, 
the angular distribution of hot electrons is highly tuneable as a function of the laser intensity and polarization, with maximal emission coming in the plane of polarization.\par


\section{\label{sec:level1}Discussion}

Hot electron generation from Liquid jets has previously been explored under slightly different laser conditions. Reports made by Morrison\cite{ Orban2015, Morrison2015} et. al. show an enhancement in the electron energies on laser interactions with liquid jets. Though the pulse energy and jet characteristics remain similar, a diffraction-limited focus takes the peak intensity up to $10^{18}$ W/cm$^2$, ideally a relativistic intensity. In their case, electrons well above 1 MeV were observed. The results surpassed the reports by Uhlrig\cite{Uhlig2011} et al., where electrons up to 300 keV were observed experimentally from water jets over an extensive intensity range. The results presented here were taken at much lower intensities ($< 10^{17}$). While higher laser intensities with tighter focusing are achievable, the low intensities utilized in this study provides greater flexibility in the Laser’s basic parameters. This flexibility in pulse-width and energy can substantially reduce the cost of a potential multi-particle source utilizing the methodology described in this paper. Additionally, higher repetition rate lasers can be employed to accelerate electrons to multiple MeV energies. While not the first to discuss high repetition rate, low-intensity MeV electron sources, this paper introduces the concept of using a liquid jet. Also the use of methanol and low vacuum requirements ease the complexities that tend to accompany intense laser based sources.\par 

\begin{figure}[!ht]
\centering
  \includegraphics[width= 0.9\linewidth]{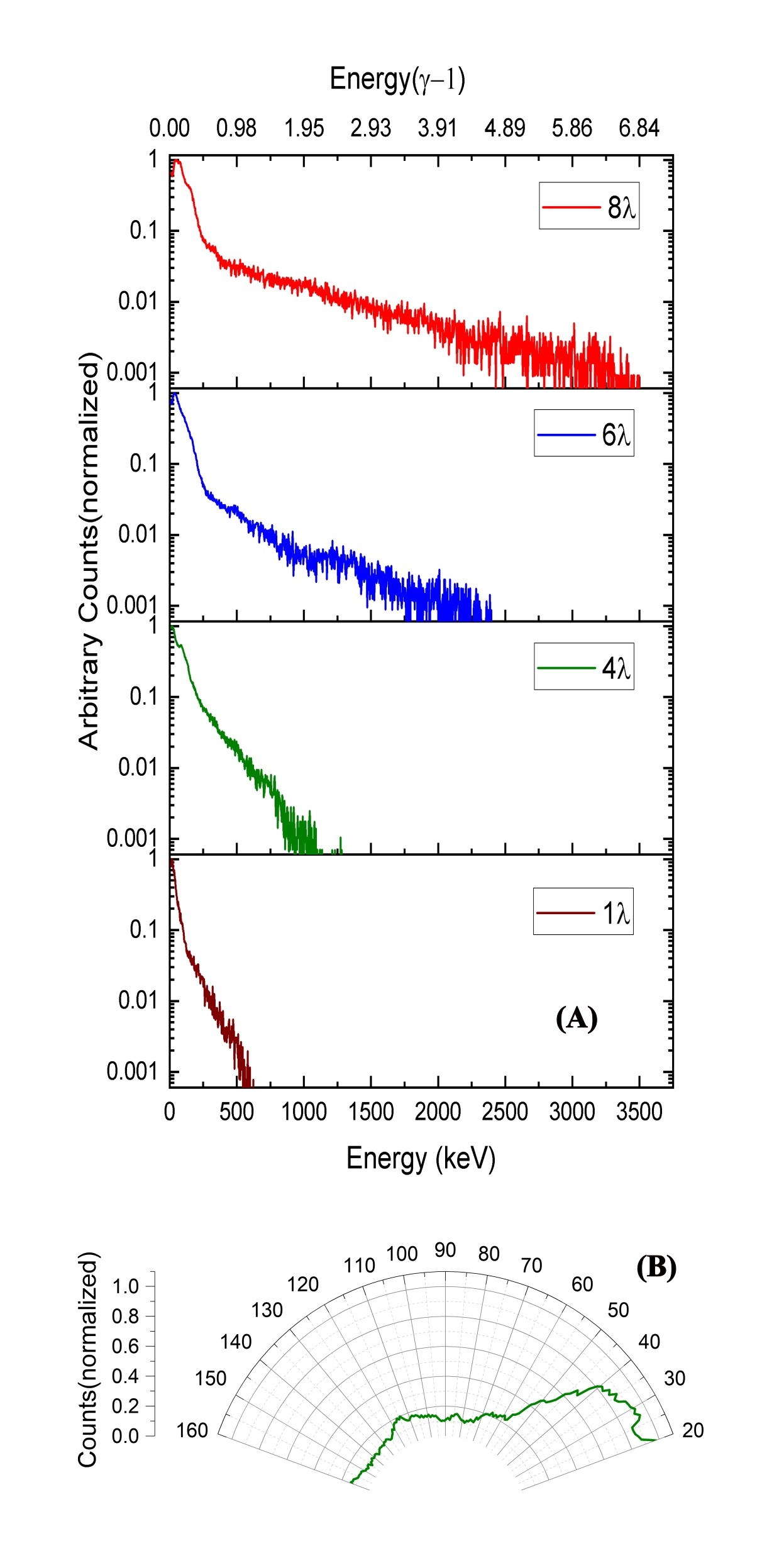}
    \caption{electron emission parameters from the PIC simulations. (A) The electron spectra recorded for the different plasma scale lengths within the hollow cup structure. The legend indicates the radially increasing plasma density gradient. (B) Angular Distribution of electrons as recorded in the PIC simulations. Here the $-\vec{k}$ of the incident laser is along the $0^\circ$ of the polar plot.    
    }
    \label{PIC_elec_total}
\end{figure}

It is important to note that electron temperatures close to 1MeV were previously possible\cite{Faure2004, Kneip2010, Lundh2011} only at intensities well above $10^{18}$W/cm$^2$, close to 100 times the value used here. To provide more perspective to the stated enhancement, a comparison can be made to the existing temperature scaling laws. Exploring scenarios\cite{Rusby2024} with and without a pre-plasma scale length provided for high $a_0$ (where $a_{0} = eE/{\omega m_{e}c}$) while aligning with ponderomotive scaling at lower ($a_{0}<1$) limits. In our case, the pondermotive scaling predicts a temperature much below 50keV, which is 20 times lower than what is experimentally measured. Though longer scale lengths predict higher temperatures, the 1 MeV component remained inexplicable even under these considerations. \par

In our earlier publications, the reasons for the anamolous high  electron temperatures was unraveled using a model where the droplet was distorted from its spherical form due to the impact of the prepulse(same as shown in Figure \ref{expSet})\cite{Sabui2024}. It was experimentally observed that apart from the expected isotropic spherical expansion, two cup-like tentacles emerged from the spherical droplet 4ns after interacting with the prepulse. The main pulse interacts with the plasma held within the concave cavity between the tentacles to produce the high-temperature component of the electron spectra. 2-dimensional radiation-hydrodynamic simulations also supported the formation of the cup structure. Various possibilities were explored to identify the optimal plasma density profile capable of generating high-energy electrons through the interaction with a moderate-intensity laser pulse. It was realized that the concave cup structure, when assumed to contain a radially increasing plasma density, could excite parametric instabilities in the sub-critical plasma layer, leading to an enhanced energy transfer. The simulations were conducted in two dimensions; thus, ideally, they should be valid both for the droplet and the liquid jet as long as the spherical symmetry is presented to the incident EM fields in the plane of polarization. \par
\begin{figure}[!ht]
\centering
  \includegraphics[width= 0.9\linewidth]{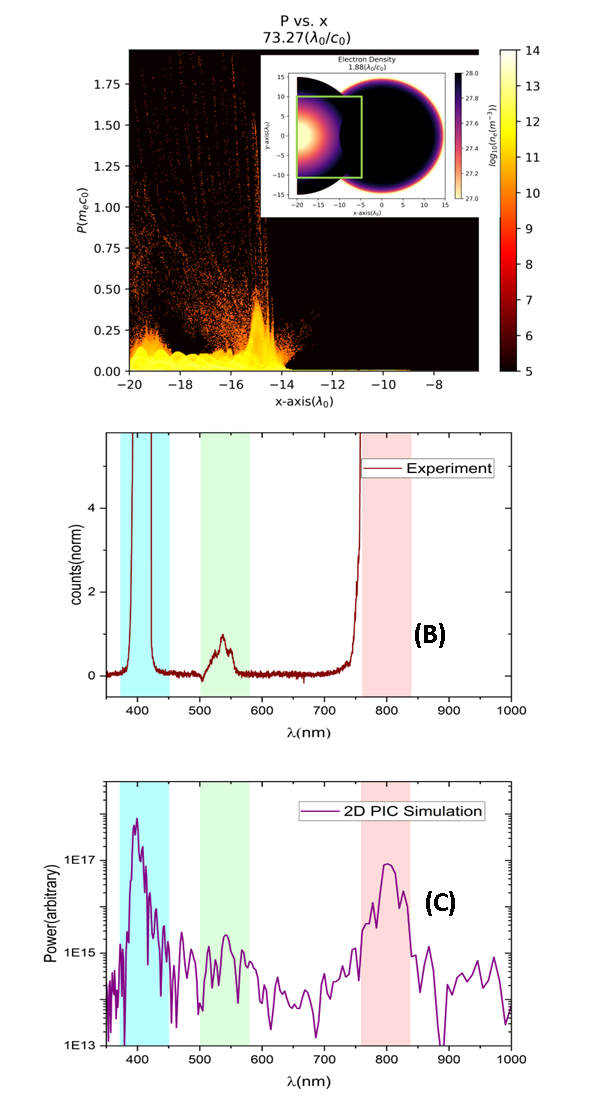}
    \caption{ (A) The plot shows the acceleration of electrons in the low density region inside the concave region of the plasma structure. Very little electron acceleration is observed within the spherical bulk plasma. For reference, the N$_c$ plasma surface is located close to -14$\lambda_0$ with the low density plasma extending from $14\lambda_0$ to $-20\lambda_0$. The N$_c$/4 region falls between $-18\lambda_0$ to $-20\lambda_0$. The inset shows the plasma density profile used in the model. The boxed region is the area containing the particles used to create the momentum-vs-space histogram shown in the main image. (B) Optical emission spectrum from the experiment involving the liquid jet. (C) Optical spectrum derived from the PIC simulations. 
    }
    \label{optspec_PICmod}
\end{figure}

Thus, a similar approach is taken for our current work, which resorts to 2 dimensional PIC simulations. The simulations were conducted using the relativistic Particle in cell code Epoch(Details in supplementary). The simulation plane represents the laser's polarization plane in 2D. The estimation was adopted as the acceleration was significant in this plane, as shown by the angular distribution. Also, the hot electron generation mechanism, as presented \cite{Mondal2024, Yembadi2024} previously under different contexts, is field-driven. The hydrodynamic expansion and resulting bulk acceleration at these pulse energies are negligible when acceleration to velocities above 10 keV is considered. Also, at the time scales being considered, this occurrence would have no effect on the spectra presented here. The plasma target is designed as a sphere with two tentacles extending in the $-\vec k$ direction, forming a cup-like structure filled with radially increasing exponential plasma density(density profile shown in Figure \ref{optspec_PICmod}(A)). \par  

The nature of electron acceleration, as realized from the 2D PIC simulations, has been shown in Figure \ref{PIC_elec_total} and \ref{optspec_PICmod}(A). The presence of the cup structure, coupled with the plasma gradient inside its concave, successfully reproduces the observed electron energies. Also, the emission angular distribution is similar to that observed in the experiments involving both the liquid droplets and the jet. A small range of plasma scale lengths was examined for their role in the acceleration of electrons. All the datasets corresponding to scale lengths beyond 4$\lambda$(shown in the figure) display a 200keV temperature component. As the plasma gradient decreases (the scale length increases), a prominent 1MeV spectral component emerges at 8$\lambda$. The 8$\lambda$ scale length was also predicted by radiation hydrodynamic simulations conducted with appropriate parameters to predict the effect of the pre-pulse. The dependence of electron energies on the subcritical plasma layer gives specific indications of the absorption mechanism at play within the system.  While simple Resonance absorption alone has previously\cite{Kluge2011, Li2003} been observed to produce electron temperatures 10$\times$ lower than our experimental results, parametric instabilities like two-plasmon decay and SRS provide a different possibility\cite{Ebrahim1980}. The incident laser EM oscillation at $\omega_0$ decomposes into Langmuir waves and less energetic EM waves at frequencies $\omega_{0}/2$ near the N$_c$/4 plasma density region. The dephasing of these Langmuir waves emits electrons at their phase velocity. A k-space analysis\cite{Mondal2024, Yembadi2024} of the fields within our concave cavity does indeed show the presence of spectral components capable of generating the experimentally detected hot electrons. The plots also show the growth of plasmons along the electron emission directions. Though these instabilities are observed for longer pulse-width lasers, steep gradients can support the localized growth of such instabilities. The presence of obliquely angled plasma gradients(due to the radial nature of the concavity) supplements the fact that our intensities are well above the TPD and SRS thresholds. \par
A secondary indication of the occurrence of TPD is the emission of 3$\omega_0$/2 radiation. Though multiple mechanisms have been proposed to support this occurrence, A simplistic explanation involves the direct coupling of the half-frequency plasmons with the pump wave ($\omega_0$ /2 with $\omega_0$) that gives rise to this effect\cite{Liu1976}. The process has also been portrayed as the scattering of the incident pump by a backward propagating plasmon. A small $\beta$ factor where $\beta =$1.41$\times$10$^{14}$T$_{e}^{2}$(keV)/[I(W/cm$^2$)$\lambda^2$]<<1 allows us imply this effect without further complications\cite{Veisz2002}. 3$\omega_0$/2 radiation was detected in the simulations and the experiments involving both the liquid droplets and the jet as shown in Figure\ref{optspec_PICmod}. At short time scales presented by our pulse, ion oscillations are not expected to play a role in the emission mechanisms.\par
While a 3D model would be optimal for understanding the precise plasma dynamics, the consistency of the outcomes of the 2D PIC model with the experiments(both jet and droplet) suggests that the phenomena at play might not differ significantly. Differences in the electron emission characteristics can be expected when the underlying mechanism involves the bulk plasma dynamics of the target. We can ignore the role of the high-density bulk in hot electron generation especially when dealing with such low intensities. The similarity in mechanism between the two targets can only occur if the formation of tentacles(due to the prepulse) observed in the droplet also happens in the jet. As the bulk mass of the droplet is very different from the jet, the bulk spherical expansion is not expected to be the same. This is another factor why the comparison of lower energy electrons becomes tricky for the system. However, the tentacle-like expansion, emerging primarily from the surface exposed to the pre-pulse, is likely to remain consistent across the 2 geometries. This is supported by the simulations by Feister et. al.\cite{Feister2016}. Since the high-temperature emission occurs only from the concavity and its contents(refer to Figure \ref{optspec_PICmod}), it remains the same for the liquid drop and the jet. Such may not be the case for extremely high Laser intensities as the relativistic effects impact denser regions of the plasma. Minor deviations in the structure play little role in the emission as has been verified by changing the structural parameters in the model.

\section{Conclusion}

In conclusion, our research highlights the feasibility of utilizing a liquid jet interaction as a potential replacement for droplets as targets in the generation of directed high temperature electrons. Droplets exploit their size-limited nature to undergo dynamic target structuring with the prepulse, resulting in an enhanced electron temperature of ~1MeV. The cylindrical geometry of the liquid jet displays its size-limited nature only along the plane of polarization, yet the acceleration characteristics are very similar to that of the 3D structure of the droplet. 2D PIC simulations, backed by experiments, were used to recreate the experimental results. The role of parametric instabilities(already established for the droplets) in electron acceleration was discussed in the context of the 2D cylindrical target. Thus we present a highly regenerative and convenient design for a directed MeV electron source using a moderate intensity Laser.

\nocite{*}
\bibliography{aipsamp}

\end{document}